\documentclass[12pt]{article}
\usepackage{amssymb}
\def\lpar{\left({\vrule height20pt width0pt depth20pt}\right.}
\def\rpar{\left.{\vrule height20pt width0pt depth20pt}\right)}
\def\dbarra{\left/{\vrule height20pt width0pt depth20pt}\right.}
\begin{document}
\baselineskip=0.5cm

\title{\bf Streams and Lazy Evaluation Applied to Integrable Models}

\author{
J.C. Brunelli$$\thanks{Email: brunelli@fsc.ufsc.br\,.}\\
{\normalsize \it Departamento de F\'\i sica, CFM, Universidade Federal de Santa Catarina,}\\
\noalign{\vspace{-.1truecm}}
{\normalsize \it Campus Universit\'ario, Trindade, C.P. 476, CEP 88040-900, Florian\'opolis, SC, Brazil}\\
 }

\maketitle

\begin{abstract}

Computer algebra procedures to manipulate pseudo-differential operators are implemented to perform calculations with integrable models. We use lazy evaluation and streams to represent and operate with pseudo-differential operators. No order of truncation is needed since terms are produced on demand. We give a series of concrete examples using the computer algebra language MAPLE. 
\end{abstract}
\bigskip
\vfill\eject

\section{Introduction}
\label{sec:introduction}

In many calculations in physics we are faced with the problem of obtaining the product of infinite power series such as
\begin{equation}
F=\sum_{i=0}^\infty f_i x^i\;,\quad G=\sum_{j=0}^\infty g_j x^j\;.\label{series}
\end {equation}
This is usually done through the convolution formula
\begin{equation}
F\star G=\sum_{i=0}^\infty x^i\sum_{j=0}^{j=i} f_j\, g_{i-j}\;.\label{seriesproduct}
\end{equation}
Implementing these products of series in a computer algebra system is usually done keeping track of indexes and expansion order of truncation, which is very cumbersome. The algebraic language MAPLE provides the package {\tt powseries} which allows us to manipulate formal power series without the need to specify the truncation order. We can ask as many terms as desired. Encapsulated in this package is a powerful paradigm of modern computation, {\em streams with delayed} or {\em lazy evaluation}, or simply {\em streams} for short. Streams are natural data structures to implement programs which use infinite objects such as sequences, series, etc. \cite{abelson} and, as we will see, pseudo-differential operators (PDO). The use of streams frees us of the task of taking care of expasion order of truncation and avoids the proliferation of summation indexes.

PDO are widely  used in the theory of integrable models \cite{das,dickey,blaszak}. Calculations with them are systematic, such as obtaining equations of motion via Lax pairs as well the generation of conserved charges via fractional powers of the Lax operators. Therefore, these calculations can be performed by the assistance of some sort of computer algebra system. In fact, a powerful program\footnote{See also \cite{ito} for other programs dealing with integrable models.} aimed to this task was introduced in \cite{seiler} using the language AXIOM \cite{axiom} (formerly known as Scratchpad). AXIOM was used because stream is an primitive data structure in this language and the existence of a large library of ``domains'' allow us to construct very generic algorithms. Unfortunately, AXIOM is not a widespread language and is not commercially available any more (efforts are being made to make AXIOM a public domain software, see {\tt http://www.nongnu.org/axiom/}). In this paper 
we will provide such algorithm in MAPLE, however, we will have to construct streams with lazy evaluation form scratch. This is in fact very straightforward \cite{abelson,gruntz}\footnote{Let us point out that the paper \cite{gruntz} was written for a version of MAPLE where nested lexical scoping was not available and therefore it had to be simulated.} and in doing so we hope to make these beautiful and elegant techniques more popular to physicists (see \cite{jerzy2} for a nice apology on these techniques).

At this point we would like to make some general comments. 
{\em Laziness}  evaluates procedure arguments only when the procedure uses them, otherwise the evaluation is delayed. This make possible to manipulate infinite data structures, such as the series (\ref{series}) and PDOs in a simple way. This technique of evaluation is the key ingredient of our algorithms. We will also use a {\em functional programming style} as much as possible to construct our algorithms in MAPLE. This will make our programs small and almost free of loops, special cases statements, synchronization of expansion orders, assignments, etc. In this style of programming, mathematical statements like
\begin{eqnarray*}
0!\!\!\!&=&\!\!\!1\;,\\
n!\!\!\!&=&\!\!\!n(n-1)!\,
\end{eqnarray*}
could be written, for instance in MAPLE, simply as
\begin{verbatim}
> new_fact:=n->`if`(n=0,1,n*new_fact(n-1));
\end{verbatim}
an almost literal transcription of a mathematical ``formula". This should be contrasted with a similar program written in an imperative programming style. In functional programming the code of programs is compact and elegant, easy to read and modify. In this example \verb|new_fact| is a {\em recursive} function or procedure and they have a key role in the functional style of programming. Also, \verb|new_fact| has a {\em terminal clause}, $n=0$, and we will see that thanks to laziness it will be possible to construct recursive functions  without terminal clauses, which we call {\em corecursive} functions. In the functional style of programming procedures must be {\em first class}, also called {\em higher order procedures}. That means that procedures can accept other procedures as parameters or can produce another procedure as their result. Laziness and the functional style of programming are some of the main characteristics of modern (pure) functional programming languages. Is the simultaneous use of them that makes possible to describe abstract mathematical objects such as PDO in a concrete way. Despite the fact that MAPLE is not a functional language, such as Haskell \cite{haskell}, procedures are first class and we can implement laziness evaluation and streams while taking advantage of the resources available in this powerful and widely used language. Even though programs using lazy evaluation (or programs written in functional languages in general), from a semantic point of view, are compact and have elegant code most of these programs use recursive calls or recursive defining relations that causes recomputation of terms several times.  In order to minimize this drawback we will use the MAPLE {\tt remember} option in our procedures\footnote{For self-referencial procedures we can also use a fixed point method \cite{gruntz,watt}.}, in that way a procedure stores its results in a remember table. Of course, this costs memory and beyond lower orders we can get memory overflow.

This paper is organized as follows. In sections 2 and 3 we introduce with detail the general procedures  to implement streams with lazy evaluation. We hope these can motivate readers to use these techniques in their own programs. In section 4 we review facts about PDO to be used in section 5 about integrable models. In section 6 we present our program  and its routines. A serie of concrete examples is given in section 7. We end the paper with some conclusions in section 8.

\section{Delay and Force}
\label{sec:delay and force}

We implement delayed evaluation following \cite{abelson} closely. The procedure\footnote{Also, following \cite{abelson} we adopt descriptive names for procedures, parameters, etc, trying to make programs as self-documenting as possible.} {\tt delay} packs any expression in a way that it can be evaluated later on demand. We use a MAPLE module to represent a record with two exports: \verb|delayed_object| and {\tt forced?}. In the  \verb|delayed_object| export we use the body of a procedure of no arguments as the delayed expression. The procedure {\tt force} will execute the procedure, forcing the evaluation. Also, this value will be stored and subsequent forcing of the same delayed object will return the stored value avoiding inefficiency problems in recursive programs with streams. The export {\tt forced?} keeps track if the expression was already forced. 
\begin{verbatim} 
> delay:=proc(s::uneval)
      option remember;
      local delay_record;
      delay_record:=module()
          export delayed_object,forced?;
      end module;
      delay_record:-forced?:=false;
      delay_record:-delayed_object:=proc()
          eval(s);
      end proc;
      Delay(delay_record);
  end proc:
\end{verbatim}
\begin{verbatim}
> force:=proc(s)
      option remember;
      local delay_record;
      delay_record:=op(1,s);
      if not delay_record:-forced? then 
          delay_record:-delayed_object:=delay_record:-delayed_object();
          delay_record:-forced?:=true;
      end if;
      delay_record:-delayed_object;
  end proc:
\end{verbatim}
\begin{verbatim}
> `print/Delay` := proc()
      "Delayed";
  end proc:
\end{verbatim}
As an simple example consider
\begin{verbatim}
> de:=delay(diff(x^2,x));
\end{verbatim}
\[
\mathit{de} := \mbox{``Delayed''}
\]
\begin{verbatim}
> eval(op([1,1,2],de));
\end{verbatim}
\[
\mathit{false}
\]
\begin{verbatim}
> force(de);
\end{verbatim}
\[
2\,x
\]
\begin{verbatim}
> eval(op([1,1,2],de));
\end{verbatim}
\[
\mathit{true}
\]

Usually, the arguments of a procedure are evaluated before the procedure is invoked. This is known as {\em applicative--order evaluation}. This behavior will return an error in the following procedure
\begin{verbatim}
> ab:=proc(a,b)
      `if`(a=0,1,b);
  end proc:
> ab(0,1/sin(0));
\end{verbatim}
\begin{flushleft}
\mbox{\tt Error, numeric exception: division by zero}
\end{flushleft}
Using {\tt delay} we can postpone the evaluation of the argument of a procedure until it is needed. This is called {\em normal-order evaluation} or {\em lazy evaluation}. In this way
\begin{verbatim}
> ab(0,delay(1/sin(0)));
\end{verbatim}
\[
1
\]
\begin{verbatim}
> ab(1,delay(1/sin(0)));
\end{verbatim}
\[
\mbox{``Delayed''}
\]
and no error is produced as long as we do not force the evaluation. This example shows us that we can perform useful computations even when values of some parameters would produce errors or are not known. We will exploit lazy evaluation in the next section to construct infinite structures.

\section{Streams}
\label{sec:streams}

Stream is a data structure representing a sequence of infinite terms. We implement streams as lists with delayed evaluation. Following the usual LISP terminology we introduce the constructor \verb|cons_stream| and two selectors, \verb|car_stream| and \verb|cdr_stream| \cite{abelson}. The procedure \verb|cons_stream| takes the two arguments {\tt head} and {\tt tail} and returns a compound object, a delayed list with the arguments as its parts
\begin{verbatim}
> cons_stream:=proc(head::uneval, tail::uneval)
      delay([head,tail]);
  end proc:
\end{verbatim}
With \verb|cons_stream| we can form pairs whose elements are pairs, and so on, i.e. we can construct infinite lists. Given a stream we can extract its head or first element using the selector \verb|car_stream|
\begin{verbatim}
> car_stream:=proc(s)
      op(1,force(s));
  end proc:
\end{verbatim}
The selector \verb|cdr_stream| selects the sublist consisting of all but the first item
\begin{verbatim}
> cdr_stream:=proc(s)
      op(2,force(s));
  end proc:
\end{verbatim}
Noticing that nested applications of \verb|car_stream| and \verb|cdr_stream}| allows us to extract any element of a stream we introduce the procedure \verb|ref_stream| which takes as arguments a stream $s$ and a number $n$ and returns the  item of the stream. It is customary to number the elements of the stream beginning with 0. The method for computing \verb|ref_stream| is the following: for $n=0$, \verb|ref_stream| should return the head of the stream, otherwise, it should return the $(n-1)$ item of the tail of the stream. Therefore, we get the recursive procedure
\begin{verbatim}
> ref_stream:=proc(s,n)
      if n=0 then
          car_stream(s);
      else
          ref_stream(cdr_stream(s),n-1);
      end if;
  end proc:
\end{verbatim}
Now we can construct our first infinite structure, a infinite list of ones
\begin{verbatim}
> stream_of_ones:=cons_stream(1,stream_of_ones);
\end{verbatim}
\[
\mathit{stream\_of\_ones} := \mbox{``Delayed''}
\]
The stream \verb|stream_of_ones| creates a list whose head equals 1 
\begin{verbatim}
> car_stream(stream_of_ones);
\end{verbatim}
\[
1
\]
but whose tail is not computed
\begin{verbatim}
> cdr_stream(stream_of_ones);
\end{verbatim}
\[
\mbox{``Delayed''}
\]
\begin{verbatim}
> car_stream(cdr_stream(stream_of_ones));
\end{verbatim}
\[
1
\]
\begin{verbatim}
> ref_stream(stream_of_ones,100);
\end{verbatim}
\[
1
\]
Note that \verb|streams_of_ones| is corecursive and thanks to the laziness the recursive evaluation is avoided. The main point to be notice with this example is that due to laziness a recursive equation has become an algorithm.

Given a function  $a(k)$ that allow us to compute the $k^{\rm th}$ element of a stream we can generate the  stream, starting from the $n^{\rm th}$ element,  $a(n+0), a(n+1), a(n+2),\dots$ by the procedure \verb|make_stream|
\begin{verbatim}
> make_stream:=proc(a,n)
      option remember;
      cons_stream(a(n),make_stream(a,n+1)):
  end proc:
\end{verbatim}
In order to display the first $m$ terms of a stream $s$ (or the $m-1$ first items  ) we use the procedure \verb|display_stream|
\begin{verbatim}
> display_stream:=proc(s,m)
      if m=0 then
          NULL;
      else
          car_stream(s),display_stream(cdr_stream(s),m-1);
      end if;
  end proc:
\end{verbatim}
Now we can represent the series (\ref{series}) as streams
\begin{verbatim}
> F:=make_stream(proc(n)
      local f;
      printf("\%a.",n);
      `if`(n=0,f[0],f[n]*x^n);
  end proc,0):
\end{verbatim}
\begin{verbatim}
> G:=make_stream(proc(n)
      local g;
      printf("\%a.",n);
      `if`(n=0,g[0],g[n]*x^n);
  end proc,0):
\end{verbatim}
Note that when a term is computed
\begin{verbatim}
> display_stream(F,10);
\end{verbatim}
\begin{flushleft}
0.1.2.3.4.5.6.7.8.9.
\end{flushleft}
\[
{f_{0}}, \,{f_{1}}\,x, \,{f_{2}}\,x^{2}, \,{f_{3}}\,x^{3}, \,{f_{
4}}\,x^{4}, \,{f_{5}}\,x^{5}, \,{f_{6}}\,x^{6}, \,{f_{7}}\,x^{7}
, \,{f_{8}}\,x^{8}, \,{f_{9}}\,x^{9}
\]
it is never recomputed
\begin{verbatim}
> display_stream(F,11);
\end{verbatim}
\begin{flushleft}
10.
\end{flushleft}
\[
{f_{0}}, \,{f_{1}}\,x, \,{f_{2}}\,x^{2}, \,{f_{3}}\,x^{3}, \,{f_{
4}}\,x^{4}, \,{f_{5}}\,x^{5}, \,{f_{6}}\,x^{6}, \,{f_{7}}\,x^{7}
, \,{f_{8}}\,x^{8}, \,{f_{9}}\,x^{9}, \,{f_{10}}\,x^{10}
\]

Let us represent the infinite series in (\ref{series}) as streams in the form of the head $f_0$ compounded with the tail power series ${\overline F}$ by \verb|cons_stream|, which we represent by ``$\,:\,$"
\[
F\equiv f_0:{\overline F}\;.
\]
The sum of the series $F$ and $G$ can then be represented by
\[
F\oplus G=(f_0:{\overline F})\oplus(g_0:{\overline G})=(f_0+g_0):({\overline F}\oplus{\overline G})\;.
\]
The product of a series $F$ by a constant $c$ is simply
\[
c\cdot F=c\cdot(f_0:{\overline F})=(cf_0):(c\cdot{\overline F})\;.
\]
Finally, the product of the series $F$ and $G$, as in (\ref{seriesproduct}), can be written as
\[
F\star G=(f_0:{\overline F})\star(g_0:{\overline G})=(f_0g_0):\left[(f_0\cdot{\overline G})\oplus({\overline F}\star G)\right]\;.
\]
We can code in MAPLE all these algorithms easily as
\begin{verbatim}
> add_stream:=proc(s1,s2)      
      cons_stream(car_stream(s1)+car_stream(s2),
          add_stream(cdr_stream(s1),cdr_stream(s2))
      );
  end proc:
\end{verbatim}
\begin{verbatim}
> display_stream(add_stream(F,G),10);
\end{verbatim}
\begin{flushleft}
\lefteqn{0.1.2.3.4.5.6.7.8.9.}
\end{flushleft}
\vspace{-1.4truecm}
\begin{eqnarray*}
&&\!\!\!\!{f_{0}} + {g_{0}}, \,{f_{1}}\,x + {g_{1}}\,x, \,{f_{2}}\,x^{2} + 
{g_{2}}\,x^{2}, \,{f_{3}}\,x^{3} + {g_{3}}\,x^{3}, \,{f_{4}}\,x^{
4} + {g_{4}}\,x^{4}, \,{f_{5}}\,x^{5} + {g_{5}}\,x^{5},\hspace{8truecm}\\
&&\mbox{}{f_{6}}\,x^{6} + {g_{6}}\,x^{6},
{f_{7}}\,x^{7} + {g_{7}}\,x^{7}, \,{f_{8}}\,x^{8} + {g_{8}}\,x^{8
}, \,{f_{9}}\,x^{9} + {g_{9}}\,x^{9}
\end{eqnarray*}
\begin{verbatim}
> scalar_multiply_stream:=proc(constant,s)
      cons_stream(constant*car_stream(s),
          scalar_multiply_stream(constant,cdr_stream(s))
      );
  end proc:
\end{verbatim}
\begin{verbatim}
> display_stream(scalar_multiply_stream(c,F),10);
\end{verbatim}
\[
c\,{f_{0}}, \,c\,{f_{1}}\,x, \,c\,{f_{2}}\,x^{2}, \,c\,{f_{3}}\,x
^{3}, \,c\,{f_{4}}\,x^{4}, \,c\,{f_{5}}\,x^{5}, \,c\,{f_{6}}\,x^{
6}, \,c\,{f_{7}}\,x^{7}, \,c\,{f_{8}}\,x^{8}, \,c\,{f_{9}}\,x^{9}
\]
\begin{verbatim}
> multiply_stream:=proc(s1,s2)   
      cons_stream(car_stream(s1)*car_stream(s2),
          add_stream(
              scalar_multiply_stream(car_stream(s1),cdr_stream(s2)),
              multiply_stream(cdr_stream(s1),s2)
          )
      );    
  end proc:
\end{verbatim}
\begin{verbatim}
> display_stream(multiply_stream(F,G),5);
\end{verbatim}
\vspace{-1.0truecm}
\begin{eqnarray*}
&&\!\!\!\!{f_{0}}\,{g_{0}}, \,{f_{0}}\,{g_{1}}\,x + {f_{1}}\,{g_{0}}\,x, \,
{f_{0}}\,{g_{2}}\,x^{2} + {f_{1}}\,{g_{1}}\,x^{2} + {f_{2}}\,{g_{0}}\,x^{2
}, \,{f_{0}}\,{g_{3}}\,x^{3} + {f_{1}}\,{g_{2}}\,x^{3}
 + {f_{2}}\,{g_{1}}\,x^{3} + {f_{3}}\,{g_{0}}\,x^{3},  \\
&& \mbox{}{f_{0}}\,{g_{4}}\,x^{4} + {f_{1}}\,{g_{3}}\,x^{4} + {f_{2}}\,{g_{2}}\,x^{4
} + {f_{3}}\,{g_{1}}\,x^{4} + {f_{4}}{g_{0}}\,x^{4}\,
\end{eqnarray*}
Now we see how the implementation of (\ref{seriesproduct}) using streams is free of indexing and order of truncation "bureaucracy". We can go on and define the quotient, reversion and other operations on power series (see references \cite{gruntz,mcilroy1,mcilroy2,jerzy1,burge}\footnote{See especially \cite{mcilroy1} , where in a pure functional language, such as Haskell, the "music of streams is played" or \cite{mcilroy2} where the "functional pearls shine".}).

A very useful operation is to apply a transformation to each element in a stream and generate the stream of results. The procedure \verb|map_stream| takes as arguments a procedure $f$ of one argument and a stream $s$, and returns a stream of the results produced by applying the procedure to each element in the stream
\begin{verbatim}
> map_stream:=proc(f,s)
      cons_stream(f(car_stream(s)),map_stream(f,cdr_stream(s)));
  end proc:
\end{verbatim}
\begin{verbatim}
> display_stream(map_stream(term->diff(term,x),F),10);
\end{verbatim}
\[
0, \,{f_{1}}, \,2\,{f_{2}}\,x, \,3\,{f_{3}}\,x^{2}, \,4\,{f_{4}}
\,x^{3}, \,5\,{f_{5}}\,x^{4}, \,6\,{f_{6}}\,x^{5}, \,7\,{f_{7}}\,
x^{6}, \,8\,{f_{8}}\,x^{7}, \,9\,{f_{9}}\,x^{8}
\]
Similar to  \verb|map_stream| we define \verb|zip_stream| which takes $n$ streams and maps a function $f$ with $n$ arguments onto them
\begin{verbatim}
> zip_stream:=proc(f,streams)
      cons_stream(f(map(car_stream,[args[2..-1]])[]),           
          zip_stream(f,map(cdr_stream,[args[2..-1]])[])
      );
  end proc:
\end{verbatim}
\begin{verbatim}
> display_stream(zip_stream((x,y)->x^2+y^2,F,G),5);
\end{verbatim}
\[
{f_{0}}^{2} + {g_{0}}^{2}, \,{f_{1}}^{2}\,x^{2} + {g_{1}}^{2}\,x
^{2}, \,{f_{2}}^{2}\,x^{4} + {g_{2}}^{2}\,x^{4}, \,{f_{3}}^{2}\,x
^{6} + {g_{3}}^{2}\,x^{6}, \,{f_{4}}^{2}\,x^{8} + {g_{4}}^{2}\,x
^{8}
\]
The procedure \verb|filter_stream| selects some elements of a given stream $s$ forming a new stream which only contains those elements for which the given predicate {\tt predicate?} is true
\begin{verbatim}
> filter_stream:=proc(predicate?,s)
      if predicate?(car_stream(s)) then   
          cons_stream(car_stream(s),
              filter_stream(predicate?,cdr_stream(s))
          );
      else
          filter_stream(predicate?,(cdr_stream(s)));
      end if;
  end proc:
\end{verbatim}
\begin{verbatim}
> display_stream(filter_stream(term->isprime(degree(term,x)),F),10);
\end{verbatim}
\begin{flushleft}
11.12.13.14.15.16.17.18.19.20.21.22.23.24.25.26.27.28.29.30.31.
\end{flushleft}
\[
{f_{2}}\,x^{2}, \,{f_{3}}\,x^{3}, \,{f_{5}}\,x^{5}, \,{f_{7}}\,x
^{7}, \,{f_{11}}\,x^{11}, \,{f_{13}}\,x^{13}, \,{f_{17}}\,x^{17}
, \,{f_{19}}\,x^{19}, \,{f_{23}}\,x^{23}, \,{f_{29}}\,x^{29}
\]

Using procedures such as \verb|map_stream|, \verb|zip_stream| and \verb|filter_stream| we can implement algorithms in  a sort of  signal-flow structure (``signals'' that flow from one  ``process stage'' to the next) if we use streams to represent ``signals'' and stream operations to implement each of the ``process stages'' \cite{abelson}.

\section{Pseudo Differential Operators}
\label{sec:pseudo differential operators}

A {\it differential operator} is the finite sum
\begin{equation}
P=\sum_{i=0}^n P_i[u]\,\partial^i\;,\label{do}
\end{equation}
where the coefficients $P_i[u]$ are differential functions and
\[
\partial\equiv {d\ \over dx}\;,
\]
i.e., $\partial$ is the differential operator with respect to the variable $x$.
Here we are considering a ring of differential operators in one independent variable $x$ and one dependent variable $u$.

For any $i,j\ge 0$ we have
\begin{equation}
\partial^i\partial^j=\partial^{i+j}\;, \label{didj}
\end{equation}
and for a differential function $Q$ we have the Leibniz rule
\begin{equation}
\partial^n Q=\sum_{k=0}^\infty\left(%
\begin{array}{c}
  n \\
  k \\
\end{array}
\right)Q^{(k)}\partial^{n-k}\;,\label{leibniz}
\end{equation}
where
\[
Q^{(k)}\equiv{d^k Q\over dx^k}\;.
\]
These two rules allow us to define the product of any two differential operators making the space of all differential operators a noncommutative ring with the constant function 1 as the identity multiplication operator.

Since (\ref{do}) is a polynomial in the total derivative $\partial$ with differential functions as coefficients we obtain pseudo-differential operators (analogous of Laurent series) if we allow negative powers of $\partial$. In this way a formal {\it pseudo-differential operator} (PDO) is the infinite series
\begin{equation}
P=\sum_{i=-\infty}^n P_i[u]\,\partial^i\;.\label{pdo}
\end{equation}
$L$ has order $n$ if its leading coefficient is not identically zero and by convention, the zero PDO is said to have order $-\infty$.

For PDOs the relation (\ref{didj}) is still valid, however, the Leibniz rule (\ref{leibniz}) generalizes to
\begin{equation}
\partial^n Q=\sum_{k=0}^\infty
{n(n-1)\cdots(n-k+1)\over k!}\,Q^{(k)}\partial^{n-k}\;,\label{genleibniz}
\end{equation}
which is valid to any $n$ and reduces to (\ref{leibniz}) if $n\ge 0$. For $n\ge 0$ (\ref{genleibniz}) can be written as
\begin{equation}
\partial^{-n} Q=\sum_{k=0}^\infty(-1)^{k}\left(%
\begin{array}{c}
  n+k-1 \\
  k \\
\end{array}
\right)Q^{(k)}\partial^{-n-k}\;.\label{generalleibniz}
\end{equation}
Again, (\ref{didj}) and (\ref{genleibniz}) allow us to define the product of any two PDOs. However, for algorithm implementation we will use the following result \cite{gelfand1} (see also \cite{olver}, exercise 5.20). For the PDOs
\begin{eqnarray*}
P &\!\!\!=\!\!\!& P[u,\partial]=\sum_{i=-\infty}^n P_i[u]\,\partial^i\;, \nonumber\\
Q &\!\!\!=\!\!\!& Q[u,\partial]=\sum_{j=-\infty}^m Q_j[u]\,\partial^j\;,
\end{eqnarray*}
we can associate the formal Laurent series
\begin{eqnarray}
P[u,p] &\!\!\!=\!\!\!& \sum_{i=-\infty}^n P_i[u]\,p^i\;, \nonumber\\
Q[u,p] &\!\!\!=\!\!\!& \sum_{j=-\infty}^m Q_j[u]\,p^j\;,\label{formallaurent}
\end{eqnarray}
by formally substituting the operator $\partial$ for the variable $p$. The product of $P$ and $Q$,
\[
R[u,\partial]=P\cdot Q\;,
\]
is then determined by the Laurent series
\begin{equation}
R[u,p]=\sum_{i=0}^\infty{1\over i!}{\partial^i P\over\partial p^i}{\partial^i Q\over\partial x^i}\;.\label{product}
\end{equation}
Rewriting  (\ref{formallaurent}) in the form
\begin{eqnarray*}
P &\!\!\!=\!\!\!& \sum_{i=0}^\infty P_i[u,p]\;, \nonumber\\
Q &\!\!\!=\!\!\!& \sum_{j=0}^\infty Q_j[u,p]\;,
\end{eqnarray*}
the expression (\ref{product}) assumes the form
\begin{equation}
R[u,p]=\sum_{k=0}^\infty\sum_{i=0}^k\sum_{j=0}^{k-i}{1\over i!}{\partial^i P_j\over\partial p^i}{\partial^i Q_{k-i-j}\over\partial x^i}\;.\label{pseudoproduct}
\end{equation}
As an example consider
\begin{eqnarray*}
P &\!\!\!=\!\!\!& \partial+u^2\partial^{-1}\;, \nonumber\\
Q &\!\!\!=\!\!\!& \partial^2+u\partial^{-1}\;,\label{examplepq}
\end{eqnarray*}
then
\begin{eqnarray*}
P[u,p] &\!\!\!=\!\!\!& p+u^2p^{-1}\;, \nonumber\\
Q[u,p] &\!\!\!=\!\!\!& p^2+u p^{-1}\;,
\end{eqnarray*}
and (\ref{pseudoproduct}) yields
\begin{eqnarray*}
R[u,p]&\!\!\!=\!\!\!&P_0Q_0+P_0Q_1+P_1Q_0+P_0Q_2+P_1Q_1+P_2Q_0\nonumber\\
&&+{\partial P_0\over\partial p}{\partial Q_1\over\partial x}+{\partial P_1\over\partial p}{\partial Q_0\over\partial x}+{\partial^2 P_0\over\partial p^2}{\partial^2 Q_0\over\partial x^2}+\cdots\nonumber\\
&\!\!\!=\!\!\!&p^3+u+u^2p+u^3p^{-2}+u_xp^{-1}-u^2u_xp^{-3}+u^2u_{xx}p^{-4}+\cdots\;,
\end{eqnarray*}
and therefore
\[
R[u,\partial]=\partial^3+u^2\partial+u+u_x\partial^{-1}+u^3\partial^{-2}-u^2u_x\partial^{-3}
+u^2u_{xx}\partial^{-4}+\cdots\;.
\]

It can be shown that every nonzero PDO $P$ has an inverse and every PDO of order $n>0$, with coefficient $P_n$, has an $n^{\rm th}$ root
\begin{equation}
N\equiv\sqrt[n]{P}=\sqrt[n]{P_n}\,\partial+a_0+a_{-1}\partial^{-1}+a_{-2}
\partial^{-2}+a_{-3}\partial^{-3}+\cdots\;.\label{root}
\end{equation}
Imposing
\[
(N)^n=P\;,
\]
we obtain a linear system of equations for the coefficients $a_n$ of $N$ which can be recursively solved. Also, fractional powers 
\begin{equation}
P^{m/n}=(\sqrt[n]{P})^m\label{mnroot}
\end{equation}
can be defined. As an example, for the operator
\begin{equation}
L=\partial^2+u\;,\label{kdvexample}
\end{equation}
we have
\begin{eqnarray}
L^{1/2}&\!\!\!\!=\!\!\!\!&\partial+{1\over 2}u\,\partial^{-1}-{1\over 4}u_x\,\partial^{-2}+{1\over 8}(u_{xx}-u^2)\,\partial^{-3}+\cdots\;,\nonumber\\
L^{3/2}&\!\!\!\!=\!\!\!\!&\partial^3+{3\over 2}u\,\partial+{3\over4}u_x+{1\over8}(3u^2+u_xx)\,\partial^{-1}-{1\over 16}(6uu_x+u_{xxx})\,\partial^{-2}
+\cdots\;.\label{rootsexample}
\end{eqnarray}
\section{Integrable Models}
\label{sec:integrable models}

Integrable systems have been studied extensively from various points of view \cite{das,dickey,blaszak} after the discovery of the inverse scattering transform. Now it is well known that integrable systems of the form
\[
u_t=K[u]\;,
\]
where $u=u(x,t)$ and $K$ is a differential function of $u$, can be represented in the form of the so called Lax equation \cite{lax,gelfand1,gelfand2}
\begin{equation}
{\partial L\over\partial t}=[B,L]\;,\label{lax}
\end{equation}
where the Lax pair $L$ and $B$ are in general PDO with scalar coefficient functions. For instance, the generalized KdV hierarchy is described in terms of the Lax pair of order $n$ \cite{gelfand3,adler,wilson,drinfeld}
\begin{eqnarray}
L_n &\!\!\!\!=\!\!\!\!& \partial^n+u_{-1}\partial^{n-1}+u_0\partial^{n-2}+\cdots+u_{n-2}\;,\nonumber\\
\noalign{\vspace{5pt}}%
B &\!\!\!\!=\!\!\!\!& \left(L_n^{k/n}\right)_+\,,\quad k\not =n\;,\label{gpairlax}
\end{eqnarray}
where $u_0(x,t),\cdots,u_{n-3}(x,t),u_{n-2}(x,t)$ represent the dynamical variables of the system and $\left(\ \right)_+$ denotes the part of the PDO with nonnegative powers. In more general cases (nonstandard Lax representation) we can have
\begin{equation}
B=\left(\ \right)_{\ge d}\;,\label{projection}
\end{equation}
representing the part of a PDO with terms $\partial^n$, $n\ge d$. In this way $\left(L_n^{k/n}\right)_+$ and $\left(L_n^{k/n}\right)_{\ge 0}$ are equivalent.
The evolution equations, or the $k^{\rm th}$ flow of the hierarchy, are given by
\begin{equation}
{\partial L_n\over\partial t_k}=[\left(L_n^{k/n}\right)_+,L_n]\;.\label{glax}
\end{equation}

The scalar Lax equation (\ref{lax}) is the compatibility condition for the system of linear equations
\begin{eqnarray*}
L\psi &\!\!\!\!=\!\!\!\!& \lambda\psi\;,\\\nonumber
{\partial\psi\over\partial t} &\!\!\!\!=\!\!\!\!& B\psi\;,
\end{eqnarray*}
imposing the constancy in time of the spectral parameter $\lambda$.

The KdV hierarchy is obtained from (\ref{gpairlax}) setting  $n=2$  and the KdV equation corresponds to $k=3$ (with $u_{-1}=0$, $u\equiv u_0$). Therefore (see (\ref{rootsexample}))
\begin{eqnarray}
L &\!\!\!\!=\!\!\!\!& \partial^2+u\;,\nonumber\\
\noalign{\vspace{5pt}}%
B &\!\!\!\!=\!\!\!\!& \left(L^{3/2}\right)_+=\partial^3+{3\over4}(\partial u+u \partial)\;,\label{kdvpair}
\end{eqnarray}
and the Lax equation (\ref{lax}) or (\ref{glax}) yields the KdV equation
\begin{equation}
u_t={1\over 4}u_{xxx}+{3\over 2} uu_x\;.\label{kdv}
\end{equation}

Given a PDO $P=\sum_{i=-\infty}^n P_i\partial_i$ its residue is defined as the coefficient of $\partial^{-1}$,
\begin{equation}
\hbox{Res}\,P=P_{-1}\;,\label{residue}
\end{equation}
and its trace is defined as
\[
\hbox{Tr}\,P=\int dx\,\hbox{Res}\,P=\int dx\,P_{-1}\;.
\]
It is easy to show that $\hbox{Tr}$ satisfies the cyclicity property \cite{adler}
\[
\hbox{Tr}\,PQ=\hbox{Tr}\,QP\;,
\]
which implies
\begin{equation}
\hbox{Tr}\,[P,Q]=0,\label{trc}
\end{equation}
for any two PDO $P$ and $Q$.

From the Lax equation (\ref{glax}) we can write
\begin{equation}
{\partial L_n^{m/n}\over\partial t_k}=[\left(L_n^{k/n}\right)_+,L_n^{m/n}]\;,\label{mnlax}
\end{equation}
for arbitrary $m$. Now, it can be shown using (\ref{glax}) and (\ref{mnlax}) that distinct flows commute \cite{wilson,drinfeld}
\[
{\partial^2\over\partial t_m\partial t_k}L_n={\partial^2\over\partial t_k\partial t_m}L_n\;.
\]
Also, taking the trace of (\ref{mnlax}) and using (\ref{trc}) we get
\[
{\partial\ \over\partial t_k}\hbox{Tr}\left(L_n^{m/n}\right)=0\;.
\]
Thus, we define the conserved charges, under any flow, as
\begin{equation}
H_m=\int dx\, h_m={n\over m}\hbox{Tr}\left(L_n^{m/n}\right)\;,\quad m=0,1,2,\dots\;.\label{charges}
\end{equation}
where $h_m={(n/ m)}\,\hbox{Res}\,(L_n^{m/n})$ is called the conserved density of the charge $H_m$.
For the KdV equation (\ref{kdv}) we have the nontrivial charges
\begin{eqnarray}
H_1 &\!\!\!\!=\!\!\!\!&\int dx\; {1\over 2}u\;,\nonumber\\
H_3 &\!\!\!\!=\!\!\!\!&\int dx\; {1\over 8}u^2\;,\nonumber\\
H_5 &\!\!\!\!=\!\!\!\!&\int dx\; {1\over 8}\left(2u^3-u_x^2\right)\;,\nonumber\\\label{kdvcharges}
&\vdots&\;.
\end{eqnarray}

Another interesting feature of integrable systems is that they are bi-Hamiltonian \cite{magri}, i.e., these systems are Hamiltonian with at least two compatible Hamiltonian structures (without loss of generality we will consider from now on the KdV hierarchy instead of the generalized KdV hierarchy (\ref{gpairlax})) 
\begin{equation}
{\partial u\over\partial t_n}=K_n[u] = \{u,H_{n+1}\}_1={\cal D}_1{\delta H_{n+1}\over\delta u}=\{u,H_{n}\}_2={\cal D}_2{\delta H_{n}\over\delta u}\;.\label{biha}
\end{equation}
Then, it can be shown that the charges (\ref{charges}) are in involution with both Hamiltonian structures 
\[
\{H_n,H_m\}_1=\{H_n,H_m\}_2=0\;,\quad n,m=0,1,2,\dots\;.
\]

For the KdV equation we have
\[
u_t={\cal D}_1{\delta H_{1}\over\delta u}={\cal D}_2{\delta H_{0}\over\delta u}\;,
\]
with
\begin{eqnarray}
{\cal D}_1&\!\!\!\!=\!\!\!\!&\partial\nonumber\;,\\
{\cal D}_2&\!\!\!\!=\!\!\!\!&\partial^3+2(u\partial+\partial u)\;.\label{bikdv}
\end{eqnarray}

Introducing the recursion operator
\[
R=\mathcal{D}_2\mathcal{D}_1^{-1}  \;,
\]
it follows from (\ref{biha}) that the hierarchy of equations can be generated by
\[
K_{n+1}=RK_n\;.
\]
Also, using
\[
R^\dagger=\mathcal{D}_1^{-1}\mathcal{D}_2\;,
\]
where $R^\dagger$ is the adjoint of $R$, a recursion scheme for the Hamiltonians $H_n$ can be written as
\[
{\delta H_{n+1}\over\delta u}=R^\dagger{\delta H_{n}\over\delta
u}\;.
\]
For a bi-Hamiltonian system of evolution equations, $u_t=K_n[u]$, a natural Lax description is given by \cite{brunelli}
\[
{\partial R\over\partial t_n}=[R,B_n]\;,
\]
where, we can identify
\[
B_n=K'_n\;.
\]
Here $K'_n$ represents the Fr\'echet derivative of $K_n$, defined
by
\[
K'_n[u]\,v=\frac{d\ }{d\epsilon}\,K_n[u+\epsilon
v]\Big|_{\epsilon=0}\Big.\;.
\]
It can also be shown that
\[
B_n\propto\left(R^{n/2}\right)_+\;,
\]
and the conserved charges are given by
\begin{equation}
H_{2n+1}\propto\hbox{Tr}\left(R^{2n+1\over2}\right)\;,\quad n=0,1,2,\dots\;.\label{rkdvcharges}
\end{equation}
For the KdV equation (\ref{kdv}) with bi-Hamiltonian structures (\ref{bikdv}) we have
\begin{eqnarray}
R&\!\!\!\!=\!\!\!\!&\mathcal{D}_2\mathcal{D}_1^{-1}=\partial^2+2u+2\partial u\partial^{-1}\;,\nonumber\\
B&\!\!\!\!=\!\!\!\!&{1\over 4}\partial(\partial^2+6u)\;,\label{rkdv}
\end{eqnarray}
and the conserved charges (\ref{kdvcharges}) also follows from (\ref{rkdvcharges}).
\section{The Program}
\label{sec:the program}

In this section we describe the procedures  in the program PSEUDO used to perform the calculations with PDO  in integrable models. In the next section we present explicit examples. Besides the procedures already introduced in section \ref{sec:delay and force} and \ref{sec:streams} to handle streams we have the following procedures to handle the PDO streams:
\bigskip

\noindent
{\underline\it Procedure name:} {\tt variables}
\smallskip\hrule height1pt
\medskip

\noindent
{\it Feature:} The procedure {\tt variables} defines the dependent variables, i.e., the evolution fields and the name of independent variables, avoiding the argument list in the input and output. 

\medskip
\noindent
{\it Calling sequence:}
\begin{verbatim}
> variables([u1,u2,..],[x,t1,t2,...]);
\end{verbatim}
\medskip

\noindent
{\it Parameters:} 
\medskip

\noindent
\begin{tabular}{lll}
\verb-u1,u2,...- & - name of the dependent variables or evolution fields
\\
\verb-x,t1,t2 ...-  & - name of the independent variables
\end{tabular}
\medskip

\noindent
{\it Examples:}
\begin{verbatim}
> read "pseudo.mpl";
> variables([u,v],[x,t]);
\end{verbatim}
\[
u, \,v
\]

\bigskip
\noindent
{\it Procedure name:} \verb|switch_diff|
\smallskip\hrule height1pt
\medskip

\noindent
{\it Feature:} Helps the visualization of derivatives and reduces the size of formulas. When activated all derivatives, for instance, of $u=u(x,t)$  will be displayed as $u_x,u_{xx},u_{xxx}$ and so on. This procedure is adapted from a similar one in the MAPLE package PDEtools \cite{cheb-terrab}.

\medskip
\noindent
{\it Calling sequence:}
\begin{verbatim}
> switch_diff(key);
\end{verbatim}
\medskip
\noindent
{\it Parameters:} 

\medskip
\noindent
\begin{tabular}{lll}
\verb-key- & - the string {\tt "on"} to switch the pretty derivative on and {\tt "off"} to turn it off.
\end{tabular}
\medskip

\noindent
{\it Examples:}
\begin{verbatim}
> expr:=diff(u,x$3)/(diff(u,x)+diff(v,x$2))^2;
\end{verbatim}
\[
\mathit{expr} := {\displaystyle \frac {{\frac {\displaystyle \partial ^{3}}{
\displaystyle \partial x^{3}}}\,u}{\left(\left({\frac {\displaystyle \partial }{\displaystyle \partial x}}\,u\right) + \left(
{\frac {\displaystyle \partial ^{2}}{\displaystyle \partial x^{2}}}\,v\right)\right)^{2}}} 
\]
\begin{verbatim}
> switch_diff("on");
\end{verbatim}
\[
\mbox{``on''}
\]
\begin{verbatim}
> expr;
\end{verbatim}
\[
{\displaystyle \frac {{u_{\mathit{xxx}}}}{({u_{x}} + {v_{\mathit{
xx}}})^{2}}} 
\]
\begin{verbatim}
> diff(expr,x);
\end{verbatim}
\[
{\displaystyle \frac {{u_{\mathit{xxxx}}}}{({u_{x}} + {v_{
\mathit{xx}}})^{2}}}  - {\displaystyle \frac {2\,{u_{\mathit{xxx}
}}\,({u_{\mathit{xx}}} + {v_{\mathit{xxx}}})}{({u_{x}} + {v_{
\mathit{xx}}})^{3}}} 
\]

We implement PDO, such as $P$ in (\ref{pdo}), as streams with lazy evaluation. In this way, no truncation order needs to be given, terms are automatically produced on demand. The terms of the PDO $P$ are represented as a sequence stream\footnote{In \cite{seiler} PDO are represented as a coefficient stream.}, starting with the term with highest degree,
\[
P_n\partial^n,\;P_{n-1}\partial^{n-1},\;P_{n-2}\partial^{n-2},\;\dots\;.
\]

\bigskip
\noindent
{\it Procedure name:} \verb|print_pseudo|
\smallskip\hrule height1pt
\medskip

\noindent
{\it Feature:}  Prints the PDO stream.
\medskip

\noindent
{\it Calling sequence:}
\begin{verbatim}
> print_pseudo(p,n);
> print_pseudo(p);
> print_pseudo(p,[n]);
\end{verbatim}
\noindent
{\it Parameters:} 

\medskip
\noindent
\begin{tabular}{lll}
\verb-p- & - a stream representing the PDO to be printed
\\
\verb-n- & - a integer corresponding to the number of  terms to be displayed in the sum
\end{tabular}

\medskip
\noindent
{\it Observation:} \verb|print_pseudo(p,n)| prints the stream, just for visualization, as a PDO with $n$ terms. If just the parameter {\tt p} is passed to the procedure then the number of printed terms will be the value set to the global variable {\tt Order}. \verb|print_pseudo(p,[n])| prints, for further manipulation, the first $n$ terms of the stream as a function.

\bigskip
\noindent
{\it Procedure name:} {\tt d}
\smallskip\hrule height1pt
\medskip

\noindent
{\it Feature:}  Creates PDOs given by the Leibniz rules (\ref{leibniz}) and (\ref{generalleibniz}). It returns a delayed stream.

\medskip
\noindent
{\it Calling sequence:}
\begin{verbatim}
> d(n,f);
> d(n);
> d(f);
> d([n],f);
\end{verbatim}
\noindent
{\it Parameters:} 

\medskip
\noindent
\begin{tabular}{lll}
\verb-n- & - a positive or negative integer representing the order of the PDO to be created
\\
\verb-f- & - a function
\end{tabular}

\medskip
\noindent
{\it Observation:} \verb|d(n,f)| stands for $\partial^n\! f$. {\tt d} also accepts one argument. If {\tt n} is absent then {\tt d(f)} represents the function $f$ as a stream and if {\tt f} is absent then {\tt d(n)} represents the pure PDO $\partial^n$ as a stream. \verb|d([n],f)| stands for $(\partial^n\! f)$, i.e., if $n\ge 0$ it represents as a stream the function 
\[
(\partial^n\! f)\equiv f_{\underbrace{\scriptstyle xx\dots x}_{\scriptstyle n}}\;,
\]
otherwise, for $n<0$, it represents\footnote{We mean an algebraic representation and not an analytical representation. The last one is given by nonlocal integrals.} as a stream the integral
\[
(\partial^n\! f)\equiv\int\int\dots\int f\, \underbrace{dxdx\dots dx}_{-n}\;.
\]
The procedure {\tt d} prints a sequence of integers as terms are forced by the first time (we will not show them in the output of the examples that follows). 
\medskip

\noindent
{\it Examples:}

\begin{verbatim}
> d(3,u);
\end{verbatim}
\[
\mbox{``Delayed''}
\]
\begin{verbatim}
> print_pseudo(%);
\end{verbatim}
\[
u\,d^{{\,{3}}} + 3\,{u_{x}}\,d^{{\,{2}}} + 3\,{u_{\mathit{xx}}}
\,d^{{\,{\ }}} + {u_{\mathit{xxx}}} + \mathit{...}
\]
We can also use \verb|display_stream| to visualize a stream
\begin{verbatim}
> display_stream(d(3,u),20);
\end{verbatim}
\[
d^{\,3}\,u, \,3\,d^{\,2}\,{u_{x}}, \,3\,{u_{\mathit{xx}}}\,d, \,{u_{
\mathit{xxx}}}, \,0, \,0, \,0, \,0, \,0, \,0, \,0, \,0, \,0, \,0
, \,0, \,0, \,0, \,0, \,0, \,0
\]
However, \verb|display_stream| does not order {\tt d}, in fact {\tt d} in any PDO output expression is always meant to be at the most right position. This example also shows us that we do not make distinction between infinite and finite PDO when we represent them as streams. \begin{verbatim}
> print_pseudo(d(-1,v),3);
\end{verbatim}
\[
v\,d^{{\,{-1}}} - {v_{x}}\,d^{{\,{-2}}} + {v_{\mathit{xx}}}\,d
^{{\,{-3}}} + \mathit{...}
\]
\begin{verbatim}
> Order:=5:
> print_pseudo(d(-1,v));
\end{verbatim}
\[
v\,d^{{\,{-1}}} - {v_{x}}\,d^{{\,{-2}}} + {v_{\mathit{xx}}}\,d
^{{\,{-3}}} - {v_{\mathit{xxx}}}\,d^{{\,{-4}}} + {v_{\mathit{
xxxx}}}\,d^{{\,{-5}}} + \mathit{...}
\]
\begin{verbatim}
> print_pseudo(d(-3));
\end{verbatim}
\[
d^{{\,{-3}}} + \mathit{...}
\]
\begin{verbatim}
> print_pseudo(d(u^2));
\end{verbatim}
\[
u^{2} + \mathit{...}
\]
\begin{verbatim}
> print_pseudo(d(u^2),100);
\end{verbatim}
\[
u^{2} + \mathit{...}
\]
\begin{verbatim}
> display_stream(d(u^2),10);
\end{verbatim}
\[
u^{2}, \,0, \,0, \,0, \,0, \,0, \,0, \,0, \,0, \,0
\]
\begin{verbatim}
> print_pseudo(d([-2],u));
\end{verbatim}
\[
{\displaystyle \int } {\displaystyle \int } u\,dx\,dx + \mathit{
...}
\]
\begin{verbatim}
> II:=print_pseudo(d([-2],u),[1]);
\end{verbatim}
\[
\mathit{II} := {\displaystyle \int } {\displaystyle \int } u\,dx
\,dx
\]
\begin{verbatim}
> print_pseudo(d([2],u));
\end{verbatim}
\[
{u_{\mathit{xx}}} + \mathit{...}
\]
\begin{verbatim}
> DD:=print_pseudo(d([2],u),[1]);
\end{verbatim}
\[
\mathit{DD} := {u_{\mathit{xx}}}
\]
\begin{verbatim}
> print_pseudo(d([1],II));
\end{verbatim}
\[
{\displaystyle \int } u\,dx + \mathit{...}
\]
\begin{verbatim}
> print_pseudo(d([-1],DD));
\end{verbatim}
\[
{u_{x}} + \mathit{...}
\]
With the procedure $\tt d$ and the following ones we can construct any PDO.

\bigskip
\noindent
{\it Procedures names:} \verb|multiply_pseudo| or {\tt \&*}, \verb|add_pseudo| or {\tt \&+}, \verb|subtract_pseudo| or {\tt \&-}
\smallskip\hrule height1pt
\medskip

\noindent
{\it Feature:}  Multiplies, adds or subtracts any number of PDOs  returning a delayed stream.

\medskip
\noindent
{\it Calling sequence:}
\begin{verbatim}
> multiply_pseudo(p1,p2,p3,...);
> &*(p1,p2);
> p1 &* p2;
> add_pseudo(p1,p2,p3,...);
> &+(p1,p2);
> p1 &+ p2;
> subtract_pseudo(p1,p2,p3,...);
> &-(p1,p2);
> p1 &- p2;
\end{verbatim}
\noindent
{\it Parameters:} 

\medskip
\noindent
\begin{tabular}{lll}
\verb-p1,p2,p3,...- & - the streams representing the PDOs to be multiplied, added or subtracted
\end{tabular}

\bigskip
\noindent
{\it Procedure name:} \verb|scalar_multiply_pseudo| or {\tt \&.}
\smallskip\hrule height1pt
\medskip

\noindent
{\it Feature:}  Multiply a PDO by  any number of functions  returning a delayed stream.

\medskip
\noindent
{\it Calling sequence:}
\begin{verbatim}
> scalar_multiply_pseudo(f1,f2,...,p);
> &.(f,p);
> f &. p;
\end{verbatim}
\noindent
{\it Parameters:} 

\medskip
\noindent
\begin{tabular}{lll}
\verb-f1,f2,...; f- & - the functions multiplying the PDO 
\\
\verb-p- & - a stream representing the PDO
\end{tabular}

\medskip
\noindent
{\it Observation:} For just two arguments the procedures {\tt \&*}, {\tt\&+}, {\tt\&-} and {\tt\&.} can be used as an unary prefix operator or as an infix binary operator. We must be very careful and observe the order of precedence of the {\tt \&} operators when used as binary operators, they are all left-associative and {\tt \&*} has the lowest binding strength: {\tt\&+}, {\tt\&-}, {\tt\&.}, {\tt\&*}. Parentheses should be used to avoid dubious inputs. Also, there is no operator overloading, therefore, care must be taken in order to not mix streams with functions.
\medskip

\noindent
{\it Examples:}
\begin{verbatim}
> print_pseudo((v^2 &. d(1,u)));
\end{verbatim}
\begin{flushleft}
\verb|Error, (in force) `delay_record` does not evaluate to a module|
\end{flushleft}
\begin{verbatim}
> print_pseudo(((v^2) &. d(1,u)));
\end{verbatim}
\[
v^{2}\,u\,d^{{\,{\ }}} + v^{2}\,{u_{x}} + \mathit{...}
\]
\begin{verbatim}
> print_pseudo(-5 &. d(2));
\end{verbatim}
\begin{flushleft}
\verb|Error, (in force) `delay_record` does not evaluate to a module|
\end{flushleft}
\begin{verbatim}
> print_pseudo((-5) &. d(2));
\end{verbatim}
\[
 - 5\,d^{{\,{2}}} + \mathit{...}
\]
\begin{verbatim}
> print_pseudo( d(1) &* d(1) &+ d(2) &- (2 &. d(2)));
\end{verbatim}
\[
 - d^{{\,{3}}} + d^{{\,{2}}} + \mathit{...}
\]
\begin{verbatim}
> print_pseudo( (d(1) &* d(1)) &+ d(2) &-  (2 &. d(2)));
\end{verbatim}
\[
\mathit{zero} + \mathit{...}
\]
\noindent For the example given by equations (\ref{examplepq})
\begin{verbatim}
> P:=d(1) &+ ((u^2) &. d(-1)):
> Q:=d(2) &+ (u &. d(-1)):
> print_pseudo(P &* Q,8);
\end{verbatim}
\[
d^{{\,{3}}} + u^{2}\,d^{{\,{\ }}} + u + {u_{x}}\,d^{{\,{-1}}}
 + u^{3}\,d^{{\,{-2}}} - u^{2}\,{u_{x}}\,d^{{\,{-3}}} + u^{2}\,
{u_{\mathit{xx}}}\,d^{{\,{-4}}} + \mathit{...}
\]

The \verb|multiply_pseudo| procedure uses the strategy described in Section \ref{sec:pseudo differential operators}. This is because implementing the product of PDO's as we have done for power series in Section \ref{sec:streams} is not straightforward. Terms of PDO acting on terms of another PDO produce another PDO, i.e, a infinite stream instead of a function. The other algebraic procedures are slightly variations of the operations implemented in Section \ref{sec:streams} for power series. 

\bigskip
\noindent
{\it Procedure name:} \verb|negate_pseudo|
\smallskip\hrule height1pt
\medskip

\noindent
{\it Feature:}  Multiplies all terms of a PDO by $-1$.

\medskip
\noindent
{\it Calling sequence:}
\begin{verbatim}
> negate_pseudo(p)
\end{verbatim}
\noindent
{\it Parameters:} 

\medskip
\noindent
\begin{tabular}{lll}
\verb-p- & - the stream representing the PDO 
\end{tabular}

\medskip
\noindent
{\it Observation:} \verb|negate_pseudo(p)| is a syntactic sugar for \verb|((-1) &. p)|.

\bigskip
\noindent
{\it Procedure name:} \verb|nth_root_pseudo|
\smallskip\hrule height1pt
\medskip

\noindent
{\it Feature:}  Calculates the $n^{\rm th}$ root of the PDO $P$ in (\ref{root}), the PDO $\sqrt[n]{P}$. 

\medskip
\noindent
{\it Calling sequence:}
\begin{verbatim}
> nth_root_pseudo(p,n)
\end{verbatim}
\noindent
{\it Parameters:} 

\medskip
\noindent
\begin{tabular}{lll}
\verb-p- & - the stream representing the PDO $P$ 
\\
\verb-n- & - a integer representing the root $n$ 
\end{tabular}

\medskip
\noindent
{\it Observation:} $n$ should divide the order of the PDO $P$. $n$ can be negative.
\medskip

\noindent
{\it Examples:}

\medskip
\noindent Using the operator (\ref{kdvexample}),
\begin{verbatim}
> L:=d(2) &+ d(u):
> A:=nth_root_pseudo(L,2):print_pseudo(A,6);
\end{verbatim}
\[
d^{{\,{\ }}} + {\displaystyle \frac {1}{2}} \,u\,d^{{\,{-1}}}
 - {\displaystyle \frac {1}{4}} \,{u_{x}}\,d^{{\,{-2}}} + \left( - 
{\displaystyle \frac {u^{2}}{8}}  + {\displaystyle \frac {1}{8}} 
\,{u_{\mathit{xx}}}\right)\,d^{{\,{-3}}} + \left({\displaystyle \frac {3}{8
}} \,u\,{u_{x}} - {\displaystyle \frac {1}{16}} \,{u_{\mathit{xxx
}}}\right)\,d^{{\,{-4}}} + \mathit{...}
\]
\noindent which is the square root in (\ref{rootsexample}).
\begin{verbatim}
> B:=nth_root_pseudo(L,-2):print_pseudo(B,6);
\end{verbatim}
\[
d^{{\,{-1}}} - {\displaystyle \frac {1}{2}} \,u\,d^{{\,{-3}}}
 + {\displaystyle \frac {3}{4}} \,{u_{x}}\,d^{{\,{-4}}} + \left(
{\displaystyle \frac {3\,u^{2}}{8}}  - {\displaystyle \frac {7}{8
}} \,{u_{\mathit{xx}}}\right)\,d^{{\,{-5}}} + \left({\displaystyle \frac {
15}{16}} \,{u_{\mathit{xxx}}} - {\displaystyle \frac {15}{8}} \,u
\,{u_{x}}\right)\,d^{{\,{-6}}} + \mathit{...}
\]
\begin{verbatim}
> print_pseudo(A &* A,10);
\end{verbatim}
\[
d^{{\,{2}}} + u + \mathit{...}
\]
\begin{verbatim}
> print_pseudo(A &* B,10);
\end{verbatim}
\[
1 + \mathit{...}
\]

\bigskip
\noindent
{\it Procedure name:} \verb|invert\_pseudo|
\smallskip\hrule height1pt
\medskip

\noindent
{\it Feature:}  Gives the inverse of a PDO.

\medskip
\noindent
{\it Calling sequence:}
\begin{verbatim}
> invert_pseudo(p)
\end{verbatim}
\noindent
{\it Parameters:} 

\medskip
\noindent
\begin{tabular}{lll}
\verb-p- & - the stream representing the PDO 
\end{tabular}

\medskip
\noindent
{\it Observation:} \verb|invert_pseudo(p)| is a syntactic sugar for \verb|nth_root_pseudo(p,-1)|.

\bigskip
\noindent
{\it Procedure name:} \verb|mnth_root_pseudo|
\smallskip\hrule height1pt
\medskip

\noindent
{\it Feature:}  Calculates the $m^{\rm th}$ power of the $n^{\rm th}$ root of the PDO $P$ in (\ref{mnroot}), the PDO $(\sqrt[n]{P})^m$. 

\medskip
\noindent
{\it Calling sequence:}
\begin{verbatim}
> mnth_root_pseudo(p,m,n)
\end{verbatim}
\noindent
{\it Parameters:} 

\medskip
\noindent
\begin{tabular}{lll}
\verb-p- & - the stream representing the PDO $P$ 
\\
\verb-m- & - a integer representing the power $m$ 
\\
\verb-n- & - a integer representing the root $n$ 
\end{tabular}

\medskip
\noindent
{\it Observation:} For $m=1$ it behaves like \verb|nth_root_pseudo|. $m$ and/or $n$ can be negative.

\medskip

\noindent
{\it Examples:}
\begin{verbatim}
> C:=mnth_root_pseudo(L,3,2):print_pseudo(C,6);
\end{verbatim}
\[
d^{{\,{3}}} + {\displaystyle \frac {3}{2}} \,u\,d^{{\,{\ }}} + 
{\displaystyle \frac {3}{4}} \,{u_{x}} + \left({\displaystyle \frac {3
\,u^{2}}{8}}  + {\displaystyle \frac {1}{8}} \,{u_{\mathit{xx}}}\right)
\,d^{{\,{-1}}} + \left( - {\displaystyle \frac {3}{8}} \,u\,{u_{x}}
 - {\displaystyle \frac {1}{16}} \,{u_{\mathit{xxx}}}\right)\,d^{{\,{
-2}}} + \mathit{...}
\]
\noindent which is the cubic root in (\ref{rootsexample}). 
Since \verb|nth_root_pseudo| and \verb|mnth_root_pseudo| accept negative arguments we can use these procedures to divide PDOs:

\bigskip
\noindent
{\it Procedure name:} {\tt \&/}
\smallskip\hrule height1pt
\medskip

\noindent
{\it Feature:}  Divides two PDOs returning a delayed stream.

\medskip
\noindent
{\it Calling sequence:}
\begin{verbatim}
> &/(p1,p2);
> p1 &/ p2;
\end{verbatim}
\noindent
{\it Parameters:} 

\medskip
\noindent
\begin{tabular}{lll}
\verb-p1,p2- & - the streams representing the PDOs to be divided 
\end{tabular}

\medskip
\noindent
{\it Observation:} {\tt \&/(p1,p2)} is a syntactic sugar for  \verb|(p1 &* nth_root_pseudo(p2,-1))|.
\medskip

\noindent
{\it Examples:}

\begin{verbatim}
> print_pseudo((L &/ C) &- invert_pseudo(A),10);
\end{verbatim}
\[
\mathit{zero} + \mathit{...}
\]
We do not work with inert PDOs, that means that any PDO is expanded, for instance
$\partial ^{-1}u$ yields
\begin{verbatim}
> 'd(-1,u)'=print_pseudo(d(-1,u),5);
\end{verbatim}
\[
\mathrm{d}(-1, \,u)=u\,d^{{\,{-1}}} - {u_{x}}\,d^{{\,{-2}}} + {
u_{\mathit{xx}}}\,d^{{\,{-3}}} - {u_{\mathit{xxx}}}\,d^{{\,{-4}
}} + {u_{\mathit{xxxx}}}\,d^{{\,{-5}}} + \mathit{...}
\]
Explicit formulae for the coefficients of PDO resulting from some operation, such as
\[
{1\over \partial u+u\partial}={1\over 2}u^{-1/2}\partial^{-1} u^{-1/2}\;,
\]
are not easy to obtain in general and are avoided in our program. However, we can still check the validity  of this equality. For the operator $OP=\partial u+u\partial$

\begin{verbatim}
> OP:=d(1,u) &+ (u &. d(1)):
> OP_invert:=invert_pseudo(OP):
> print_pseudo(OP_invert,3);
\end{verbatim}
\[
{\displaystyle \frac {1}{2}} \,{\displaystyle \frac {d^{{\,{-1}}
}}{u}}  + {\displaystyle \frac {1}{4}} \,{\displaystyle \frac {{u
_{x}}\,d^{{\,{-2}}}}{u^{2}}}  - {\displaystyle \frac {1}{8}} \,
{\displaystyle \frac {( - 3\,{u_{x}}^{2} + 2\,u\,{u_{\mathit{xx}}
})\,d^{{\,{-3}}}}{u^{3}}}  + \mathit{...}
\]
\begin{verbatim}
> print_pseudo(OP_invert &* OP,10);
\end{verbatim}
\[
1 + \mathit{...}
\]
\begin{verbatim}
> OP_invert_exact:=(u^(-1/2)/2) &. d(-1,u^(-1/2)):
> print_pseudo(OP_invert_exact &* OP,10);
\end{verbatim}
\[
1 + \mathit{...}
\]
\begin{verbatim}
> print_pseudo(OP_invert_exact &- OP_invert,10);
\end{verbatim}
\[
\mathit{zero} + \mathit{...}
\]

\bigskip
\noindent
{\it Procedure name:} \verb|residue_pseudo|
\smallskip\hrule height1pt
\medskip

\noindent
{\it Feature:}  Gives the residue (\ref{residue}) of the PDO.

\medskip
\noindent
{\it Calling sequence:}
\begin{verbatim}
> residue_pseudo(p);
\end{verbatim}
\noindent
{\it Parameters:} 

\medskip
\noindent
\begin{tabular}{lll}
\verb-p- & - a stream representing the PDO
\end{tabular}

\medskip

\noindent
{\it Examples:}
\begin{verbatim}
> residue_pseudo(C);
\end{verbatim}
\[
{\displaystyle \frac {3\,u^{2}}{8}}  + {\displaystyle \frac {1}{8
}} \,{u_{\mathit{xx}}}
\]

\bigskip
\noindent
{\it Procedure name:} \verb|project_pseudo|
\smallskip\hrule height1pt
\medskip

\noindent
{\it Feature:}  Returns the part of the PDO as indicated in (\ref{projection}). The result is returned as a stream.

\medskip
\noindent
{\it Calling sequence:}
\begin{verbatim}
> project_pseudo(p,d)
\end{verbatim}
\noindent
{\it Parameters:} 
\medskip

\noindent
\begin{tabular}{lll}
\verb-p- & - a stream representing the PDO
\\
\verb-d- & - a integer representing the degree of the projection
\end{tabular}

\medskip

\noindent
{\it Examples:}
\begin{verbatim}
> E:=project_pseudo(C,0):print_pseudo(E);
\end{verbatim}
\[
d^{{\,{3}}} + {\displaystyle \frac {3}{2}} \,u\,d^{{\,{\ }}} + 
{\displaystyle \frac {3}{4}} \,{u_{x}} + \mathit{...}
\]

\bigskip
\noindent
{\it Procedure name:} \verb|commutator_pseudo|
\smallskip\hrule height1pt
\medskip

\noindent
{\it Feature:}  Calculates the commutator of two PDO returning a stream as the resulting PDO.

\medskip
\noindent
{\it Calling sequence:}
\begin{verbatim}
> commutator_pseudo(p1,p2)
\end{verbatim}
\noindent
{\it Parameters:} 

\medskip
\noindent
\begin{tabular}{lll}
\verb-p1,p2- & - the streams representing the PDOs
\end{tabular}

\medskip

\noindent
{\it Examples:}
\begin{verbatim}
> F:=commutator_pseudo(L,E):print_pseudo(F);
\end{verbatim}
\[
 - {\displaystyle \frac {3}{2}} \,u\,{u_{x}} - {\displaystyle 
\frac {1}{4}} \,{u_{\mathit{xxx}}} + \mathit{...}
\]

\bigskip
\noindent
{\it Procedure name:} \verb|differentiate_pseudo|
\smallskip\hrule height1pt
\medskip

\noindent
{\it Feature:}  Differentiates each term of a PDO with respect with some variable returning a stream as the resulting PDO

\medskip
\noindent
{\it Calling sequence:}
\begin{verbatim}
> differentiate_pseudo(p,x)
\end{verbatim}
\noindent
{\it Parameters:} 

\medskip
\noindent
\begin{tabular}{lll}
\verb-p- & - a stream representing the PDO to be differentiated
\\
\verb-x- & - the variable of differentiation
\end{tabular}

\bigskip
\noindent
{\it Procedure name:} \verb|equations_pseudo|
\smallskip\hrule height1pt
\medskip

\noindent
{\it Feature:}  Prints, just for visualization, a number of equations resulting in taking each coefficients of a stream and making it equal to zero. Useful to print the equations produced by Lax equations.

\medskip
\noindent
{\it Calling sequence:}
\begin{verbatim}
> equations_pseudo(p,n)
\end{verbatim}
\noindent
{\it Parameters:} 

\medskip
\noindent
\begin{tabular}{lll}
\verb-p-& - a stream representing a PDO
\\
\verb-n- & - a integer representing the number of equations to be printed
\end{tabular}

\medskip

\noindent
{\it Examples:}

\begin{verbatim}
> equations_pseudo(differentiate_pseudo(L,t) &+ F,1); 
\end{verbatim}
\[
{u_{t}} - {\displaystyle \frac {3}{2}} \,u\,{u_{x}} - 
{\displaystyle \frac {1}{4}} \,{u_{\mathit{xxx}}}=0
\]
\noindent This is of course the KdV equation (\ref{kdv}) that follows from the Lax pair of operators (\ref{kdvpair}) after using the Lax equations (\ref{lax}). 

\section{Examples}
\label{sec:examples}

In this section we present an interactive MAPLE session with the Input/Output data of some classical examples as well some results obtained recently with the aid of the program PSEUDO. We start with
\begin{verbatim}
> read "pseudo.mpl":
> switch_diff("on"):
> Order:=10:
\end{verbatim}

\subsection{The KdV Hierarchy}

The classical KdV hierarchy of equations follows from the Lax operator (\ref{kdvexample}),

\begin{verbatim}
> variables([u],[x,t]):
> L_kdv:=d(2) &+ d(u);
\end{verbatim}
\[
\mathit{L\_kdv} := \mbox{``Delayed''}
\]
\begin{verbatim}
> print_pseudo(L_kdv);
\end{verbatim}
\[
d^{{\,{2}}} + u + \mathit{...}
\]
The square root (\ref{rootsexample}) can be calculated from
\begin{verbatim}
> print_pseudo(mnth_root_pseudo(L_kdv,1,2));
\end{verbatim}

\begin{eqnarray*}
&&\!\!\!\!\!\!d^{{\,{\ }}} + {\displaystyle \frac {1}{2}} \,u\,d^{{\,{-1}}}
 - {\displaystyle \frac {1}{4}} \,{u_{x}}\,d^{{\,{-2}}} + \left( - 
{\displaystyle \frac {u^{2}}{8}}  + {\displaystyle \frac {1}{8}} 
\,{u_{\mathit{xx}}}\right)\,d^{{\,{-3}}} + \left({\displaystyle \frac {3}{8
}} \,u\,{u_{x}} - {\displaystyle \frac {1}{16}} \,{u_{\mathit{xxx
}}}\right)\,d^{{\,{-4}}} \\
&&\mbox{} + \left({\displaystyle \frac {u^{3}}{16}}  - {\displaystyle 
\frac {7}{16}} \,u\,{u_{\mathit{xx}}} - {\displaystyle \frac {11
}{32}} \,{u_{x}}^{2} + {\displaystyle \frac {1}{32}} \,{u_{
\mathit{xxxx}}}\right)\,d^{{\,{-5}}} \\
&&\mbox{} + \left( - {\displaystyle \frac {15}{32}} \,u^{2}\,{u_{x}} + 
{\displaystyle \frac {15}{32}} \,u\,{u_{\mathit{xxx}}} + 
{\displaystyle \frac {15}{16}} \,{u_{x}}\,{u_{\mathit{xx}}} - 
{\displaystyle \frac {1}{64}} \,{u_{\mathit{xxxxx}}}\right)\,d^{{\,{-6
}}}\\
&&\mbox{} + \left({\displaystyle \frac {85}{64}} \,u\,{u_{x}}^{2} + 
{\displaystyle \frac {55}{64}} \,u^{2}\,{u_{\mathit{xx}}} - 
{\displaystyle \frac {31}{64}} \,u\,{u_{\mathit{xxxx}}} - 
{\displaystyle \frac {37}{32}} \,{u_{x}}\,{u_{\mathit{xxx}}} + 
{\displaystyle \frac {1}{128}} \,{u_{\mathit{xxxxxx}}} - 
{\displaystyle \frac {91}{128}} \,{u_{\mathit{xx}}}^{2} - 
{\displaystyle \frac {5\,u^{4}}{128}} \right)\,d^{{\,{-7}}}\\
&&\mbox{} + \left( - {\displaystyle \frac {175}{32}} \,u\,{u_{x}}\,{u_{
\mathit{xx}}} + {\displaystyle \frac {175}{128}} \,{u_{x}}\,{u_{
\mathit{xxxx}}} + {\displaystyle \frac {35}{64}} \,u^{3}\,{u_{x}}
 - {\displaystyle \frac {175}{128}} \,u^{2}\,{u_{\mathit{xxx}}}
 - {\displaystyle \frac {1}{256}} \,{u_{\mathit{xxxxxxx}}} - 
{\displaystyle \frac {175}{128}} \,{u_{x}}^{3} \right.\\
&&\mbox{}\left. + {\displaystyle \frac {245}{128}} \,{u_{\mathit{xx}}}\,{
u_{\mathit{xxx}}} + {\displaystyle \frac {63}{128}} \,u\,{u_{
\mathit{xxxxx}}}\right) d^{{\,{-8}}}\mbox{} + \mathit{...}
\end{eqnarray*}
The Lax equation
\[
{\partial L_{\rm KdV}\over\partial t}=[(L_{\rm KdV}^{n/2})_+,L_{\rm KdV}]\;,
\]
will produce nontrivial equations for $n$ odd. The first equations are
\begin{verbatim}
> for n from 1 to 8 by 2 do equations_pseudo(differentiate_pseudo(L_kdv,t)
  &+ commutator_pseudo(L_kdv,project_pseudo(mnth_root_pseudo(L_kdv,n,2),0))
  ,10) end do;
\end{verbatim}

\[
{u_{t}} - {u_{x}}=0
\]
\[{u_{t}} - {\displaystyle \frac {3}{2}} \,u\,{u_{x}} - 
{\displaystyle \frac {1}{4}} \,{u_{\mathit{xxx}}}=0
\]
\[
 - {\displaystyle \frac {15}{8}} \,u^{2}\,{u_{x}} - 
{\displaystyle \frac {5}{8}} \,u\,{u_{\mathit{xxx}}} - 
{\displaystyle \frac {5}{4}} \,{u_{x}}\,{u_{\mathit{xx}}} - 
{\displaystyle \frac {1}{16}} \,{u_{\mathit{xxxxx}}} + {u_{t}}=0
\]
\[
 - {\displaystyle \frac {35}{16}} \,u^{3}\,{u_{x}} - 
{\displaystyle \frac {35}{32}} \,u^{2}\,{u_{\mathit{xxx}}} - 
{\displaystyle \frac {35}{8}} \,u\,{u_{x}}\,{u_{\mathit{xx}}} - 
{\displaystyle \frac {7}{32}} \,u\,{u_{\mathit{xxxxx}}} - 
{\displaystyle \frac {35}{32}} \,{u_{x}}^{3} - {\displaystyle 
\frac {35}{32}} \,{u_{\mathit{xx}}}\,{u_{\mathit{xxx}}} - 
{\displaystyle \frac {21}{32}} \,{u_{x}}\,{u_{\mathit{xxxx}}} 
\]
\[
\mbox{} - {\displaystyle \frac {1}{64}} \,{u_{\mathit{xxxxxxx}}}
 + {u_{t}}=0\hspace{10truecm}
\]
The first nontrivial odd conserved densities are
\begin{verbatim}
> for n from 1 to 8 by 2 do residue_pseudo(mnth_root_pseudo(L_kdv,n,2))
  end do;
\end{verbatim}

\[
{\displaystyle \frac {u}{2}}
\]
\[
{\displaystyle \frac {3\,u^{2}}{8}}  + {\displaystyle \frac {1}{8
}} \,{u_{\mathit{xx}}} 
\]
\[
{\displaystyle \frac {5\,u^{3}}{16}}  + {\displaystyle \frac {5}{
16}} \,u\,{u_{\mathit{xx}}} + {\displaystyle \frac {5}{32}} \,{u
_{x}}^{2} + {\displaystyle \frac {1}{32}} \,{u_{\mathit{xxxx}}}
\]
\[
{\displaystyle \frac {35}{64}} \,u\,{u_{x}}^{2} + {\displaystyle 
\frac {7}{32}} \,{u_{x}}\,{u_{\mathit{xxx}}} + {\displaystyle 
\frac {35}{64}} \,u^{2}\,{u_{\mathit{xx}}} + {\displaystyle 
\frac {7}{64}} \,u\,{u_{\mathit{xxxx}}} + {\displaystyle \frac {
35\,u^{4}}{128}}  + {\displaystyle \frac {21}{128}} \,{u_{
\mathit{xx}}}^{2} + {\displaystyle \frac {1}{128}} \,{u_{\mathit{
xxxxxx}}}
\]
\medskip

\noindent
Assuming that $u\to 0$ sufficiently fast as $|x|\to\infty$ we obtain from (\ref{charges}) (after integration by parts and up to normalization constants) the conserved charges (\ref{kdvcharges}).

Alternatively, we can obtain \cite{brunelli} (up to normalization constants) all results for the KdV hierarchy using the recursion operator (\ref{rkdv})  
\begin{verbatim}
> R_kdv:=d(2) &+ d(4*u) &+ ((2*diff(u,x)) &. d(-1));
\end{verbatim}
\[
\mathit{R\_kdv} := \mbox{``Delayed''}
\]
\begin{verbatim}
> print_pseudo(R_kdv);
\end{verbatim}
\[
d^{{\,{2}}} + 4\,u + 2\,{u_{x}}\,d^{{\,{-1}}} + \mathit{...}
\]
\begin{verbatim}
> print_pseudo(mnth_root_pseudo(R_kdv,1,2));
\end{verbatim}

\begin{eqnarray*}
&&\!\!\!\!\!\!d^{{\,{\ }}}+ 2\,u\,d^{{\,{-1}}} - 2\,u^{2}\,d^{{\,{-3}}} + 4
\,u\,{u_{x}}\,d^{{\,{-4}}} + (4\,u^{3} - 2\,{u_{x}}^{2} - 4\,u\,
{u_{\mathit{xx}}})\,d^{{\,{-5}}} \\
\noalign{\vspace{3pt}}
&&\mbox{} + ( - 24\,u^{2}\,{u_{x}} + 4\,{u_{x}}\,{u_{\mathit{xx}}}
+ 4\,u\,{u_{\mathit{xxx}}})\,d^{{\,{-6}}}\\
\noalign{\vspace{3pt}}
&&\mbox{} + (52\,u\,{u_{x}}^{2} + 40\,u^{2}\,{u_{\mathit{xx}}} - 2
\,{u_{\mathit{xx}}}^{2} - 10\,u^{4} - 4\,u\,{u_{\mathit{xxxx}}}
 - 4\,{u_{x}}\,{u_{\mathit{xxx}}})\,d^{{\,{-7}}}\\
 \noalign{\vspace{3pt}}
&& \mbox{} + (4\,{u_{x}}\,{u_{\mathit{xxxx}}} - 36\,{u_{x}}^{3} - 192\,u\,{u_{
x}}\,{u_{\mathit{xx}}} + 4\,u\,{u_{\mathit{xxxxx}}} + 4\,{u_{
\mathit{xx}}}\,{u_{\mathit{xxx}}} + 120\,u^{3}\,{u_{x}} - 60\,u^{
2}\,{u_{\mathit{xxx}}}) \\
\noalign{\vspace{3pt}}
&&d^{{\,{-8}}}\mbox{} + \mathit{...}
\end{eqnarray*}

\begin{verbatim}
> for n from 1 to 8 by 2 do
  equations_pseudo(differentiate_pseudo(R_kdv,t) &+ 
  commutator_pseudo(R_kdv,project_pseudo(mnth_root_pseudo(R_kdv,n,2),0))
  ,10) end do;
\end{verbatim}
\[
4\,{u_{t}} - 4\,{u_{x}}=0 
\]
\[
2\,{u_{\mathit{tx}}} - 2\,{u_{\mathit{xx}}}=0 
\]
\vspace{.5pt}
\[
4\,{u_{t}} - 24\,u\,{u_{x}} - 4\,{u_{\mathit{xxx}}}=0
\]
\[
2\,{u_{\mathit{tx}}} - 12\,{u_{x}}^{2} - 12\,u\,{u_{\mathit{xx}}}
 - 2\,{u_{\mathit{xxxx}}}=0
 \]
\vspace{.5pt}
\[
4\,{u_{t}} - 4\,{u_{\mathit{xxxxx}}} - 120\,u^{2}\,{u_{x}} - 40\,
u\,{u_{\mathit{xxx}}} - 80\,{u_{x}}\,{u_{\mathit{xx}}}=0 
\]
\[
2\,{u_{\mathit{tx}}} - 120\,u\,{u_{x}}^{2} - 60\,{u_{x}}\,{u_{
\mathit{xxx}}} - 60\,u^{2}\,{u_{\mathit{xx}}} - 40\,{u_{\mathit{
xx}}}^{2} - 20\,u\,{u_{\mathit{xxxx}}} - 2\,{u_{\mathit{xxxxxx}}}
=0
\] 
\vspace{.5pt}
\[
4\,{u_{t}} - 560\,u^{3}\,{u_{x}} - 280\,u^{2}\,{u_{\mathit{xxx}}}
 - 1120\,u\,{u_{x}}\,{u_{\mathit{xx}}} - 56\,u\,{u_{\mathit{xxxxx
}}} - 280\,{u_{x}}^{3} - 168\,{u_{x}}\,{u_{\mathit{xxxx}}}
\]
\[
\mbox{} - 280\,{u_{\mathit{xx}}}\,{u_{\mathit{xxx}}} - 4\,{u_{
\mathit{xxxxxxx}}}=0 \hspace{9truecm}
\]
\vspace{1pt}
\[
- 840\,u^{2}\,{u_{x}}^{2} - 840\,u\,{u_{x}}\,{u_{\mathit{xxx}}}
 - 980\,{u_{x}}^{2}\,{u_{\mathit{xx}}} - 112\,{u_{x}}\,{u_{
\mathit{xxxxx}}} - 280\,u^{3}\,{u_{\mathit{xx}}} - 560\,u\,{u_{
\mathit{xx}}}^{2}
\]
\[\mbox{} - 224\,{u_{\mathit{xx}}}\,{u_{\mathit{xxxx}}} - 140\,{u_{
\mathit{xxx}}}^{2} - 140\,u^{2}\,{u_{\mathit{xxxx}}} - 28\,u\,{u
_{\mathit{xxxxxx}}} - 2\,{u_{\mathit{xxxxxxxx}}} + 2\,{u_{
\mathit{tx}}}=0
\]
\begin{verbatim}
> for n from 1 to 8 by 2 do residue_pseudo(mnth_root_pseudo(R_kdv,n,2))
  end do;
\end{verbatim}
\vspace{-.5truecm}
\[{2\,u }\]
\[{6\,u^{2} + 2\,{u_{\mathit{xx}}} }\]
\[{20\,u^{3} + 20\,u\,{u_{\mathit{xx}}} + 10\,{u_{x}}^{2} + 2\,{u_{
\mathit{xxxx}}} }\]
\[{140\,u\,{u_{x}}^{2} + 56\,{u_{x}}\,{u_{\mathit{xxx}}} + 140\,u^{2
}\,{u_{\mathit{xx}}} + 28\,u\,{u_{\mathit{xxxx}}} + 70\,u^{4} + 
42\,{u_{\mathit{xx}}}^{2} + 2\,{u_{\mathit{xxxxxx}}} }\]

\subsection{The KP Hierarchy}

In order to treat the KP hierarchy we need to construct a generic PDO, $L_{\rm KP}$. This is easily done by using the \verb|make_stream| procedure
\begin{verbatim}
> variables([U],[x,t[1],t[2],t[3]]):
> L_KP:=add_pseudo(d(1),make_stream(proc(k)
      option remember;
      U[k]*d^(-k);
  end proc,1)):
> print_pseudo(L_KP);
\end{verbatim}
\[
d^{{\,{\ }}} + {U_{1}}\,d^{{\,{-1}}} + {U_{2}}\,d^{{\,{-2}}}
 + {U_{3}}\,d^{{\,{-3}}} + {U_{4}}\,d^{{\,{-4}}} + {U_{5}}\,d^{
{\,{-5}}} + {U_{6}}\,d^{{\,{-6}}} + {U_{7}}\,d^{{\,{-7}}} + {U
_{8}}\,d^{{\,{-8}}} + \mathit{...}
\]
The Lax equation
\[
{\partial L_{\rm KP}\over\partial t_n}=[\left(L_{\rm KP}^{n}\right)_+,L_{\rm KP}]\;,\quad n=1,2,\dots\;,
\]
will produce the equations in the hierarchy. For $n=1,2,3$ we obtain
\begin{verbatim}
> equations_pseudo(differentiate_pseudo(L_KP,t[1]) &+      
  commutator_pseudo(L_KP,project_pseudo(L_KP,0)),3);
\end{verbatim}
\[
{U_{{\mathit{3t}_{1}}}} - {U_{\mathit{3x}}}=0
\]
\[
{U_{{\mathit{2t}_{1}}}} - {U_{\mathit{2x}}}=0
\]
\[
{U_{{\mathit{1t}_{1}}}} - {U_{\mathit{1x}}}=0
\]
\begin{verbatim}
> equations_pseudo(differentiate_pseudo(L_KP,t[2]) &+ 
  commutator_pseudo(L_KP,project_pseudo(L_KP &* L_KP,0)),3);
\end{verbatim}
\[
{U_{{\mathit{2t}_{2}}}} - 2\,{U_{1}}\,{U_{\mathit{1x}}} - 2\,{U_{
\mathit{3x}}} - {U_{\mathit{2xx}}}=0
\]
\[
{U_{{\mathit{1t}_{2}}}} - 2\,{U_{\mathit{2x}}} - {U_{\mathit{1xx}
}}=0
\]
\[
{U_{{\mathit{3t}_{2}}}} - 4\,{U_{2}}\,{U_{\mathit{1x}}} + 2\,{U_{
1}}\,{U_{\mathit{1xx}}} - 2\,{U_{\mathit{4x}}} - {U_{\mathit{3xx}
}}=0
\]
\begin{verbatim}
> equations_pseudo(differentiate_pseudo(L_KP,t[3]) &+ 
  commutator_pseudo(L_KP,project_pseudo(L_KP &* L_KP &* L_KP,0)),3);
\end{verbatim}
\[
{U_{{\mathit{1t}_{3}}}} - 6\,{U_{1}}\,{U_{\mathit{1x}}} - 3\,{U_{
\mathit{3x}}} - 3\,{U_{\mathit{2xx}}} - {U_{\mathit{1xxx}}}=0
\]
\[
{U_{{\mathit{2t}_{3}}}} - 6\,{U_{2}}\,{U_{\mathit{1x}}} - 6\,{U_{
1}}\,{U_{\mathit{2x}}} - 3\,{U_{\mathit{4x}}} - 3\,{U_{\mathit{
3xx}}} - {U_{\mathit{2xxx}}}=0
\]
\[
{U_{{\mathit{3t}_{3}}}} - 9\,{U_{3}}\,{U_{\mathit{1x}}} - 6\,{U_{
2}}\,{U_{\mathit{2x}}} + 3\,{U_{2}}\,{U_{\mathit{1xx}}} + 3\,{U_{
1}}\,{U_{\mathit{2xx}}} - 3\,{U_{\mathit{5x}}} - 3\,{U_{1}}\,{U_{
\mathit{3x}}} - 3\,{U_{\mathit{4xx}}} - {U_{\mathit{3xxx}}}=0
\]

\subsection{The Harry Dym Hierarchy}

The Harry Dym equation 
\[
w_t=(w^{-1/2})_{xxx}=
-{\displaystyle \frac {1}{8}} \,{\displaystyle \frac {15
\,{w_{x}}^{3} - 18\,{w_{x}}\,w\,{w_{\mathit{xx}}} + 4\,w^{2}\,{w
_{\mathit{xxx}}}}{w^{(7/2)}}} 
\label{harrydym}\;,
\]
and its hierarchy \cite{brunelli1} can be obtained from the Lax representation
\begin{eqnarray*}
L_{\rm HD}&=&{1\over w}\partial^2\;,\\
{\partial L_{\rm HD}\over\partial
t}&=&-2^n\left[\left(L_{\rm HD}^{{2n+1\over2}}\right)_{\ge2},L_{\rm HD}\right]\;,\quad n=0,1,2,\dots\;,
\end{eqnarray*}
Therefore, the first equations are
\begin{verbatim}
> `~`:=``:
> assume(w(x,t)>0):
> variables([w],[x,t]):
> L_hd:=(1/w) &. d(2):print_pseudo(L_hd);
\end{verbatim}
\[
{\displaystyle \frac {d^{{\,{2}}}}{w}}  + \mathit{...}
\]
\begin{verbatim}
> for n from 0 to 3 do equations_pseudo(differentiate_pseudo(L_hd,t) &+
  commutator_pseudo(project_pseudo(mnth_root_pseudo(L_hd,2*n+1,2),2),
  (2^n) &. L_hd),5) end do;
\end{verbatim}
\[
 - {\displaystyle \frac {{w_{t}}}{w^{2}}} =0
\]

\[
 - {\displaystyle \frac {1}{8}} \,{\displaystyle \frac {8\,{w_{t}
}\,w^{\left(\displaystyle{7\over2}\right)} + 15\,{w_{x}}^{3} - 18\,{w_{x}}\,w\,{w_{\mathit{xx}}
} + 4\,w^{2}\,{w_{\mathit{xxx}}}}{w^{\left(\displaystyle{11\over2}\right)}}} =0
\]

\[
 -{\displaystyle \frac {1}{128}}
 \lpar 128\,{w_{t}}\,w^{\left(\displaystyle{13\over2}\right)} + 
3465\,{w_{x}}^{5} - 6930\,w\,{w_{x}}^{3}\,{w_{\mathit{xx}}} + 
2520\,w^{2}\,{w_{x}}\,{w_{\mathit{xx}}}^{2} + 1820\,w^{2}\,{w_{x}
}^{2}\,{w_{\mathit{xxx}}}
\]
\vspace{-.9truecm}
\[
\hspace{-1.0truecm}\phantom{w^{\left(\displaystyle{1\over1}\right)}}\mbox{} - 560\,w^{3}\,{w_{\mathit{xx}}}\,{w_{\mathit{xxx}}} - 320
\,w^{3}\,{w_{x}}\,{w_{\mathit{xxxx}}} + 32\,w^{4}\,{w_{\mathit{
xxxxx}}}\rpar\dbarra
 \!  \! w^{\left(\displaystyle{17\over2}\right)}=0\hspace{2.5truecm}
\]

\[
 - {\displaystyle \frac {1}{4096}} \lpar 4096\,{w_{t}}\,w^{\left(\displaystyle{19\over2}\right)} + 
360192\,w^{4}\,{w_{x}}\,{w_{\mathit{xx}}}\,{w_{\mathit{xxxx}}} - 
3052896\,w^{3}\,{w_{x}}^{2}\,{w_{\mathit{xxx}}}\,{w_{\mathit{xx}}
}\hspace{2.truecm}
\]
\vspace{-.3truecm}
\[
\mbox{} + 225792\,w^{4}\,{w_{x}}\,{w_{\mathit{xxx}}}^{2} - 
10720710\,w\,{w_{x}}^{5}\,{w_{\mathit{xx}}} + 2942940\,w^{2}\,{w
_{x}}^{4}\,{w_{\mathit{xxx}}}\hspace{2.6truecm}
\]
\[
\mbox{} - 591360\,w^{3}\,{w_{x}}^{3}\,{w_{\mathit{xxxx}}}  - 32256\,w^{5}\,{w_{\mathit{xxx}}}\,{w_{\mathit{xxxx}}}
 - 21504\,w^{5}\,{w_{\mathit{xx}}}\,{w_{\mathit{xxxxx}}}\hspace{2.6truecm}
\]
\[
\mbox{} + 
310464\,w^{4}\,{w_{\mathit{xx}}}^{2}\,{w_{\mathit{xxx}}} + 87360
\,w^{4}\,{w_{x}}^{2}\,{w_{\mathit{xxxxx}}}- 8960\,
{w_{x}}\,w^{5}\,{w_{\mathit{xxxxxx}}}\hspace{2.7truecm}
\]
\[
+ 8072064\,w^{2}\,{w_{x}}^{3}\,{w_{\mathit{xx}}}^{2} - 
1397088\,w^{3}\,{w_{\mathit{xx}}}^{3}\,{w_{x}} + 512\,w^{6}\,{w_{
\mathit{xxxxxxx}}}\hspace{3.2truecm}
\]
\[
\mbox{}  + 3828825\,{w_{x}}^{7}\rpar \dbarra \!  \! 
w^{\displaystyle\left(23\over2\right)}=0\hspace{8.1truecm}
\]
\noindent
The charges follows from
\[
H_{-(n+1)}=\hbox{Tr} L_{\rm HD}^{2n-1\over 2}\;,\quad n=0,1,2,\dots\;,
\]
and the first conserved densities are
\begin{verbatim}
> for n from 0 to 3 do residue_pseudo(mnth_root_pseudo(L_hd,2*n-1,2)) 
  end do;
\end{verbatim}
\[
\sqrt{w}
\]
\[
{\displaystyle \frac {1}{32}} \,{\displaystyle \frac {5\,{w_{x}}
^{2} - 4\,w\,{w_{\mathit{xx}}}}{w^{\displaystyle\left(5\over2\right)}}} 
\]
\[
 - {\displaystyle \frac {1}{2048}} \,{\displaystyle \frac { - 
1155\,{w_{x}}^{4} + 1848\,{w_{x}}^{2}\,w\,{w_{\mathit{xx}}} - 448
\,{w_{x}}\,w^{2}\,{w_{\mathit{xxx}}} - 336\,w^{2}\,{w_{\mathit{xx
}}}^{2} + 64\,w^{3}\,{w_{\mathit{xxxx}}}}{w^{\displaystyle\left(11\over2\right)}}} 
\]
\[
 - {\displaystyle \frac {1}{65536}} ( 42944\,w^{3}\,{w_{\mathit{xx
}}}^{3} - 425425\,{w_{x}}^{6} - 569712\,{w_{x}}^{2}\,w^{2}\,{w_{
\mathit{xx}}}^{2} + 181632\,w^{3}\,{w_{\mathit{xx}}}\,{w_{
\mathit{xxx}}}\,{w_{x}}
\]
\[
\mbox{} + 52800\,{w_{x}}^{2}\,w^{3}\,{w_{\mathit{xxxx}}} - 6912\,
{w_{x}}\,w^{4}\,{w_{\mathit{xxxxx}}} + 1021020\,{w_{x}}^{4}\,w\,{
w_{\mathit{xx}}} - 8832\,w^{4}\,{w_{\mathit{xxx}}}^{2}
\]
\[
\mbox{} - 274560\,{w_{x}}^{3}\,w^{2}\,{w_{\mathit{xxx}}} + 512\,w
^{5}\,{w_{\mathit{xxxxxx}}} - 14592\,w^{4}\,{w_{\mathit{xx}}}\,{w_{\mathit{xxxx}}}) {\left/{\vrule height5pt width0pt depth5pt}\right.} \!  \! w^{\displaystyle\left(17\over2\right)}\hspace{1.5truecm}
\]
After integration by parts and up to normalization constants the conserved charges can be written as
\begin{eqnarray*}
H_{-1}&=&\int  dx\,2w^{1/2}\nonumber\;,\\\noalign{\vskip
5pt}
H_{-2}&=&\int  dx\,{1\over8}w^{-5/2}w_x^2\nonumber\;,\\\noalign{\vskip 5pt}
H_{-3}&=&\int  dx\,{1\over
16}\left({35\over16}w^{-11/2}w_x^4-w^{-7/2}w_{xx}^2\right)\nonumber\;,\\\noalign{\vskip
5pt}
H_{-4}&=&\int  dx\,{1\over
32}\left({5005\over128}w^{-17/2}w_x^6-{231\over8}w^{-13/2}w_x^2w_{xx}^2+
5w^{-11/2}w_{xx}^3\right.\nonumber\\
&&\left.\phantom{1\over2}+w^{-9/2}w_{xxx}^2\right)\nonumber\;.\\
\end{eqnarray*}
The Hunter-Zheng equation
\[
w_t=-(\partial^{-2}w)w_x-2(\partial^{-1}w)w\;,
\]
also belongs to the Harry Dym hierarchy. Its Lax representation is given by
\[
 {\partial L_{\rm HD}\over\partial
t}=-2[B,L_{\rm HD}]\;,
\]
where
\[
B={1\over 4}(\partial^{-2}w)\partial+{1\over4}\partial^{-1}(\partial^{-2}w)\partial^2\;.
\]
In fact, we can check this result
\begin{verbatim}
> B:=((1/4) &. (d([-2],w) &* d(1))) &+ ((1/4) &. (d(-1) &* d([-2],w) 
  &* d(2))):print_pseudo(B,4);
\end{verbatim}
\[
{\displaystyle \frac {1}{2}} \,{\displaystyle \int } 
{\displaystyle \int } w\,dx\,dx\,d^{{\ _{\ }}} - {\displaystyle 
\frac {1}{4}} \,{\displaystyle \int } w\,dx + {\displaystyle 
\frac {1}{4}} \,w\,d^{{\ _{-1}}} - {\displaystyle \frac {1}{4}} 
\,{w_{x}}\,d^{{\ _{-2}}} + \mathit{...}
\]
\begin{verbatim}
> equations_pseudo(differentiate_pseudo(L_hd,t) &+ 
  commutator_pseudo(scalar_multiply_pseudo(2,B),L_hd),1);
\end{verbatim}
\[
 - {\displaystyle \frac {{w_{t}} + 2\,{\displaystyle \int } w\,dx
\,w + {\displaystyle \int } {\displaystyle \int } w\,dx\,dx\,{w_{
x}}}{w^{2}}} =0
\]
\section{Conclusion}
\label{sec:Conclusion}

Using lazy evaluation and streams we have described our set of routines to perform calculations with integrable models using PDOs. In order to make the techniques used available for a wide audience we have introduced in detail the procedures to incorpore lazy evaluation as well streams in MAPLE.

Our program  provides just a basic set of operations in a way that more complicated procedures can be easily constructed. Our program just works for one dimensional PDOs . We intend to generalize this program to incorporate PDO with more general coefficients, such as matrix. A version to incorporate supersymmetric PDOs, among others generalizations, is under development. 
\section*{Acknowledgments}

We would like to thank Professor Dominik Gruntz for sending us a copy of the paper ``Infinite Structures in Maple''. Much of the work described here is based on his paper.

\end{document}